\documentclass[aps,prd,reprint,groupedaddress]{revtex4-1}

\pdfoutput=1
\usepackage{amsmath}
\usepackage{amssymb}
\usepackage{graphicx,mathrsfs}
\usepackage{nicefrac}

\newcommand{\be}{\begin{equation}}
\newcommand{\ee}{\end{equation}}
\newcommand{\bea}{\begin{eqnarray}}
\newcommand{\eea}{\end{eqnarray}}

\begin{document}


\title{Scattering Light by Light at 750 GeV at the LHC}


\author{S.~Fichet}
\affiliation{ICTP-SAIFR, IFT, S\~ao Paulo State University, Brazil}

\author{G.~von Gersdorff}

\affiliation{ICTP-SAIFR, IFT, S\~ao Paulo State University, Brazil}
\affiliation{Departamento de F\'isica, Pontif\'icia Universidade Cat\'olica de Rio de Janeiro, Rio de Janeiro, Brazil}

\author{C.~Royon}
  \affiliation{Kansas University, Lawrence, USA}
  \affiliation{Institute of Physics, Academy of Science, Prague, Czech Republic}
  \affiliation{Nuclear Physics Institute (PAN), Cracow, Poland}



\begin{abstract}

We consider the possibility that the diphoton excess at 750 GeV is caused by a new scalar resonance produced in photon fusion.
 This scenario is parametrised by only one relevant effective couplings and is thus minimal. We show that this setup can reproduce both the  production rate and width of the resonance, and is not in conflict with the 8 TeV limits on the diphoton cross section. 
The scenario also predicts event rates for $WW$, $ZZ$, $Z\gamma$ final states.  We suggest to perform precision measurements by studying light-by-light scattering with intact protons detected in forward detectors.
 We construct a simple model that shows that the required couplings can be achieved with new vectorlike, uncolored fermions (with a strong Yukawa coupling to the resonance) which may also account for the required width.


\end{abstract}


\maketitle

\section{Introduction}

Recently, the ATLAS and CMS Collaborations reported a small excess in di-photon production, in the invariant mass region between 690 and 810 GeV~\cite{CMS_note,ATLAS_note}, in the first $\mathscr{L}=$3.2 fb$^{-1}$ of the 13~Tev collisions.
 While it is too early at this stage to know if this excess is real or if it is due to statistical fluctuations,
it is important to discuss which beyond Standard Model (BSM) physics might explain this excess and how to test these hypotheses further.

Since a spin-1 resonance cannot decay into two photons by Yang's theorem, we must consider resonances of spin-0 and spin-2. 
In this letter we will focus on the spin-0 CP even case, even though the treatment of spin-0 CP odd and spin-2 cases is very similar \cite{Fichet:2015yia}.

As couplings of spin-0 states to light quarks are expected to be suppressed by the masses of the latter, one would be inclined to conclude that the best way to produce such resonances at a hadron collider is via gluon fusion \cite{Dittmaier:2011ti}. 
However, this would imply also a sizeable cross section into gluon jets, given by 
$\sigma_{gg}\sim \Gamma_{gg}/\Gamma_{\gamma\gamma}\,\sigma_{\gamma\gamma}$ in the narrow width approximation.
In a scenario where the couplings to gluons and photons are induced by new colored and charged particles, the partial width into gluons  scale as $\Gamma_{gg}/\Gamma_{\gamma\gamma}\sim\alpha_s^2/\alpha^2\sim 200$. This leads  to a dijet cross section at 13 TeV of $\sigma_{gg}\sim 1$ pb, which would have already shown up in the data. This conclusion might be avoided with some extra model-building, but one has to also keep in mind the increasingly strong exclusion limits on new colored particles from the data of run I.

In this letter, we point out an alternative possibility to gluon fusion production. Assuming that the $750$ GeV scalar resonance coupling to gluons and quarks is zero or very small,
we propose instead that the resonance could be  mostly produced in electroweak processes. The most important of these processes is vector boson fusion (collisions of $W$'s, $Z$'s and photons), which is well known from SM Higgs boson production. Unlike for the Higgs, however, a sizeable coupling of the resonance to photons is expected. This naturally leads  to the possibility that the scalar resonance be produced via \textit{photon fusion}. Such a scenario provides in fact a minimal interpretation of the diphoton excess, and is thus rather attractive.

 This letter is organised as follows. We will first estimate the effective coupling of the putative resonance to photons required to explain the excess in photon fusion.   We will  then discuss predictions for diboson final states and for  elastic light-by-light scattering at the 13~TeV LHC.
  Finally, we will present a simple UV model of vectorlike fermions that can generate the necessary couplings in loops as well as the decay width.

%


\section{Effective Parametrization of the Diphoton Excess}

\subsection{Effective Couplings}

A complete effective description of a scalar resonance (that we will denote by $\phi$) coupled to the SM has been presented in Ref.~\cite{Fichet:2015yia}. The coupling of the resonance to the photons is uniquely described by two operators
\be
\mathcal L=\frac{1}{f_B}\phi\,(B_{\mu\nu})^2+\frac{1}{f_W}\phi\,(W_{\mu\nu}^a)^2
=\frac{1}{f_{\gamma}}\phi\,(F_{\mu\nu})^2+\dots
\label{eq:OB_OW}
\ee
with $f_{\gamma}^{-1}=c_w^2\,f_B^{-1}+s_w^2\,f_W^{-1}$ with $s_w$ ($c_w$) being the sine (cosine) of the weak mixing angle. The ellipses correspond to  couplings to $ZZ$, $Z\gamma$ and $WW$.


The only other operator in the EFT of Ref.~\cite{Fichet:2015yia} involving electroweak gauge bosons is $\phi |D_\mu H|^2$. This operator does not couple $\phi$ to  photons, but can in principle contribute to its production via weak boson fusion.
However, we know from observation that $\phi$ must couple to either $(B_{\mu\nu})^2$ or $(W_{\mu\nu})^2$. It will become clear from the next sections that these operators alone can give a large enough contribute to $\phi$ production. We will therefore disregard the contributions from $\phi |D_\mu H|^2$ to $\phi$ production in this analysis \footnote{The latter operator also gives a much smaller contribution to the production for a coefficient of the same size as those of the field strength operators.}.

The partial decay width into photons induced by the operators in Eq.~(\ref{eq:OB_OW}) is
\be
\Gamma_{\gamma\gamma}=\frac{m_\phi^3}{4\pi f_{\gamma}^2}.
\ee
Of course, as is clear from the effective operators in the unbroken EW phase of Eq.~(\ref{eq:OB_OW}), a coupling to the photon is always accompanied by a coupling to $Z$, and possibly to $W$'s. The decay width induced by these operators is given by $\Gamma_{VV}=m^3(f_B^{-2}+3f_W^{-2})/4\pi$. For a fixed decay width into photons, the ratio  $\Gamma_{VV}/\Gamma_{\gamma\gamma}$ is thus constrained to be in the range
\be
\frac{3}{3c^4_w+s^4_w}\leq \frac{\Gamma_{VV}}{\Gamma_{\gamma\gamma}} \leq \frac{3}{s^4_w}\,.
\ee
For $s^2_w\approx 0.23$, these bounds are respectively $1.64$ and $56.7$.
The $\Gamma_{VV}$ width constitutes the minimal contributions to the total width $\Gamma_{\rm tot}$ of the scalar resonance. Other contributions to the width potentially exist, either induced from other operators of the EFT, for example $\phi |D_\mu H|^2$, or from decays into other BSM particles.

\subsection{The Photon Fusion Cross Section} 
 
The couplings of the scalar resonance in Eq.~(\ref{eq:OB_OW}) induce a production via  vector boson fusion diagrams. Although this process is familiar from SM Higgs boson production, a crucial difference is that it happens via the  $(B_{\mu\nu})^2$, $(W_{\mu\nu})^2$  operators, unlike for the case of the Higgs which couples mainly to $(Z_\mu)^2$, $W_\mu^+W_\mu^-$. As a result, not only the weak bosons enter in the VBF diagrams but also the photons. 

It can be readily checked at leading-order that photon fusion diagrams actually dominate the VBF cross-section.
The leading correction comes from the interference between photon and weak boson diagrams, that typically modifies the pure photon fusion cross section by $O(10\%)$. 
As the photon is massless, the photon fusion diagrams dominate because they tend to be  singular in the collinear limit.
However, a computation at leading order is a poor approximation, because large collinear logarithms should be taken into account \cite{Martin:2004dh}.  
Moreover, in the forward limit the singularities are cutoff by the  finite-size effects of the proton. This implies that  the cross-section crucially depends on the proton form factors.

\begin{figure}
\includegraphics[width=5cm]{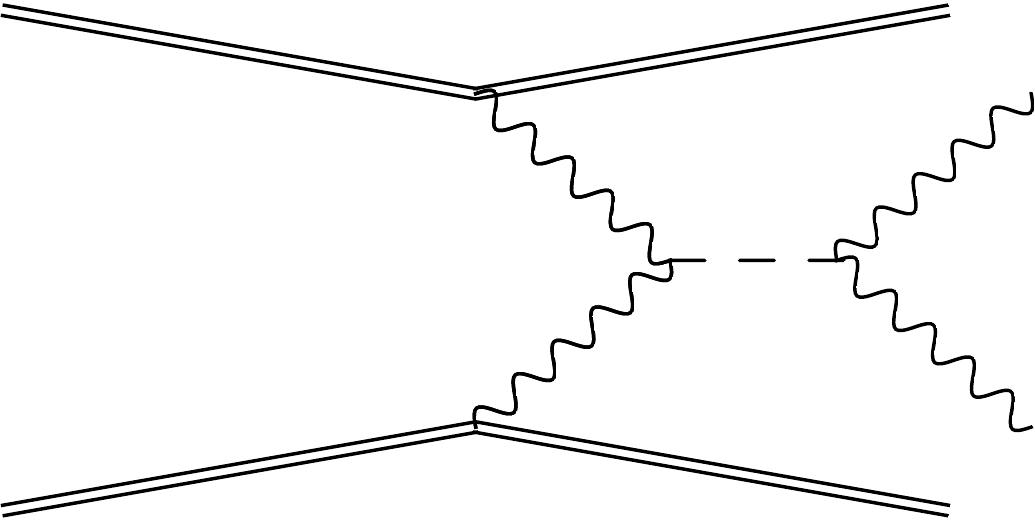}\\
\vspace{.5cm}
\includegraphics[width=5cm]{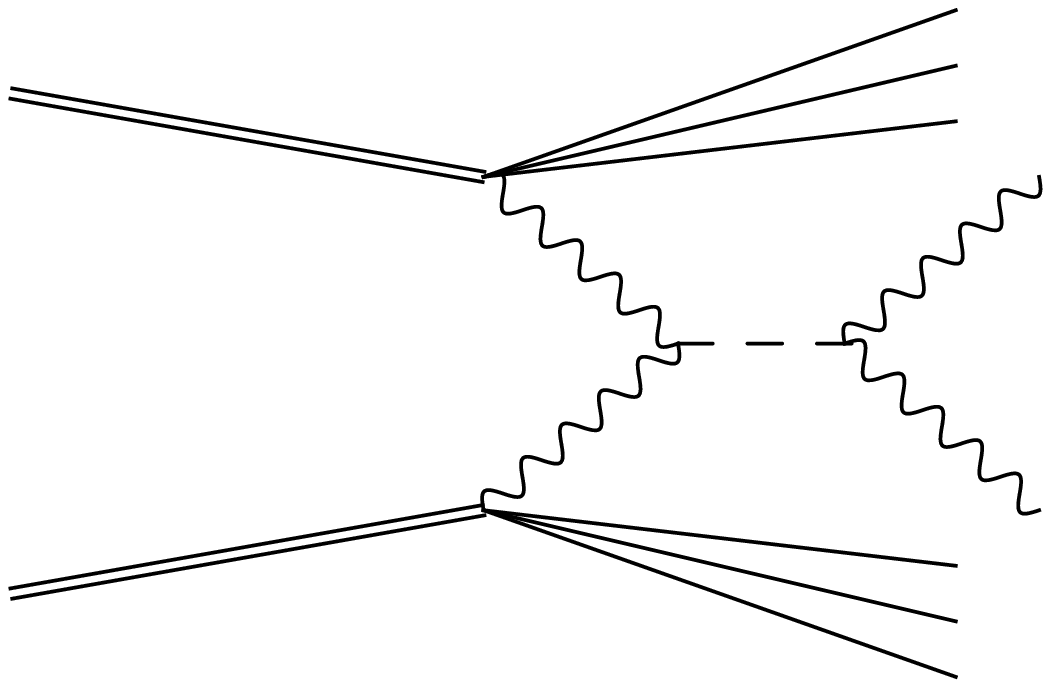}%
\caption{\label{diagrams}Schematic representation of elastic (above) and inelastic (below) photon fusion production of the resonance with  subsequent decay into photons. }
\end{figure}

Another peculiarity of photon fusion is that an emission of a photon from a proton may leave the proton intact, a process referred to as elastic scattering. This is in contrast to the inelastic case, in which the proton is destroyed. The two cases are schematically shown in Fig.~\ref{diagrams}. The inelastic case is expected to be largely dominant, whereas the elastic case is much cleaner and can in principle be identified with forward detectors. As the current diphoton searches do not distinguish the two cases, we have to assume the (dominant) inelastic case.
The various subtleties and  unknowns in the evaluation of photon fusion cross-section will be discussed below. Here we rather provide directly the result that we shall use in the rest of our analysis. The $13$~TeV cross-section is obtained to be
\be
\sigma_{pp\rightarrow \gamma\gamma X}=  ( 7.3~{\rm fb})\, \left(\frac{5\, {\rm TeV}}{f_{\gamma}}\right)^4 \left(\frac{45~{\rm GeV}}{\Gamma_{\rm tot}}\right) \left(\frac{r_{\rm inel}}{20}\right) r_{\rm fs}\,.
\label{eq:XS}
\ee

\noindent The factor $r_{\rm inel}$ denotes the ratio of the inelastic over the elastic photon fusion cross section, which is in the range 
 $r_{\rm inel}\in[15,25]$ reflecting the uncertainty 
coming from the lack of knowledge of inelastic fluxes. 
The factor $r_{\rm fs}$ is a reduction factor due to the finite size of the proton, it can be taken in the range 
 $r_{\rm fs}\in [0.4,1]$ parametrizing the uncertainty that the proton model induces on the cross-section. 
 Even though these theoretical uncertainties appear important, the determination of $f_{\gamma}$ from the data will not depend too much on them due to the fact that it enters in the fourth power in (\ref{eq:XS}).
   In the following, our baseline scenario corresponds to $\Gamma_{\rm tot}=45$~GeV, $r_{\rm fs}=1$, $r_{\rm inel}=20$.

\section{ Bounds on effective couplings and predictions for the 13~TeV LHC}

\subsection{ Implications for the Effective Photon Coupling}

The ATLAS Collaboration reported an excess of $\hat s=15$ events over a smooth diphoton invariant mass distribution,
fitted from the data over the range $m_{\gamma\gamma}=[200,1600]$~GeV.
 The excess appears in the $[690,810]$~GeV region. The expected number of background events over this range is estimated to be $b=22.7$ from the background fit.  The mass of the potential resonance  is about $750-760$~GeV,   and the total width obtained through an unbinned analysis is  $\Gamma_{\rm tot}=45$ GeV. 
This excess is compatible with the analogous search by CMS, who see a mild excess in the same mass region \cite{CMS_note}.

In our model  of a scalar resonance mainly coupled to electroweak operators, the measurement of the total rate readily provides a constraint on $f_{\gamma}$.
The efficiency $\epsilon_{\gamma\gamma}$ of the signal selection is estimated to be $45\%$ in the case of weak boson fusion. We can safely use this number for photon fusion, as this is a very similar process. 
Using the observed number of events, the expected background,  the production cross-section Eq.~(\ref{eq:XS}), and the efficiency $\epsilon_{\gamma\gamma}$, we can write down a likelihood
\be
L(f_\gamma)=P(\hat s+b|\mathscr{L}\, \sigma_{pp\rightarrow \gamma\gamma X}(f_\gamma)\epsilon_{\gamma\gamma} )\,,
\ee
where $P(N|\lambda)$ is the Poisson distribution with parameter $\lambda$. The maximum likelihood occurs at
$f_{\gamma}\approx 
4.6~{\rm TeV} $ and credible intervals can be given as 
\be
\frac{f_\gamma}{\rm TeV}\in[4.1,5.2]\ {\rm at }\ 68\%{\rm CL},\,\ [3.8,7.2]\ {\rm at }\ 95\%{\rm CL}\,,
\ee
 assuming a log prior for $f_\gamma$. For the photon fusion cross section $\sigma_{pp\rightarrow \gamma\gamma X}$, these intervals  are translated as
\be
  [8.55,\ 23.7]~{\rm fb}\ {\rm at }\ 68\%{\rm CL},\, [2.50,\ 33.2]~{\rm fb}\ {\rm at }\ 95\%{\rm CL}\,.
\ee
The constraint from the 8 TeV data on the diphoton cross section is about $\sigma_{\gamma\gamma}^{8\rm\, TeV}<3$ fb \cite{Khachatryan:2015qba,Aad:2015mna}. Ignoring the finite size effects of the proton, the photon flux is about a factor of 2.4 higher at 13 TeV. The 8 TeV constraint would thus imply that at 13 TeV we would have expected the cross section to be smaller than $\sim 7.2$ fb. Very roughly speaking this is already consistent, but we will see later that finite size effects further improve the consistency.

\subsection{Implications for the Electroweak Couplings}

\begin{figure}
\includegraphics[width=7cm]{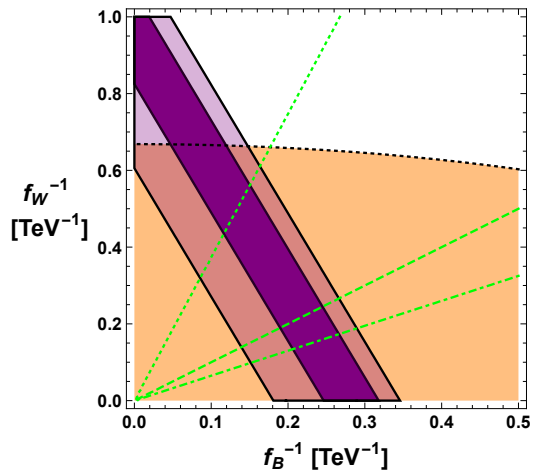}\\
\caption{\label{fig:OB_OW} Bounds in the $f_B-f_W$ plane from the event rate and the total width of the diphoton excess. The purple regions correspond to $68\%$CL and  $95\%$CL credible regions. The orange region satisfies  $\Gamma_{VV} \leq\Gamma_{\rm tot}=45$~GeV. The bound is saturated at the edge of the region.
Isolines for $\frac{{\Gamma}_{ZZ}}{{\Gamma}_{\gamma\gamma}}=1$, $\frac{{\Gamma}_{WW}}{{\Gamma}_{\gamma\gamma}}=1$, $\frac{{\Gamma}_{Z\gamma}}{{\Gamma}_{\gamma\gamma}}=1$ are respectively shown as dashed, dotdashed and dotted lines.
}
\end{figure}

This constraint from the total rate readily provides a constraint on the $f_W$, $f_B$ coefficients in Eq.~(\ref{eq:OB_OW}). Moreover, another, independent constraint comes from requiring that the partial width generated by these operator do not exceed the observed total width, \textit{i.e.}~$\Gamma_{VV}\leq \Gamma_{\rm tot}$. These two constraints are shown in Fig.~\ref{fig:OB_OW}.
It turns out that the total rates and the width are consistent over a large region of the parameter space, except when the $\phi (W_{\mu\nu}^a)^2$ operator dominates. Over the region where event rates and total width are compatible, 
  one  can further observe that the minimal contribution  to the width $\Gamma_{VV}$ is well below $\Gamma_{\rm tot}$. This implies that a substantial room is left for decays into other states besides $\gamma\gamma$, $ZZ$, $W^+W^-$, $Z\gamma$.
  
The effective theory also predicts the ratios of partial widths between the various decays $\phi\rightarrow \gamma\gamma, ZZ, Z\gamma, W^+W^-$ as a function of $f_B$ and $f_W$. Writing $r=f_W/f_B$, it follows that
\be\begin{split}
&\frac{{\Gamma}_{ZZ}}{{\Gamma}_{\gamma\gamma}}=\frac{(s^2_w r +c^2_w )^2}{(s^2_w +c^2_w r )^2}\,,\quad 
\frac{{\Gamma}_{Z\gamma}}{{\Gamma}_{\gamma\gamma}}=\frac{2c^2_ws^2_w(1-r)^2}{(s^2_w  +c^2_w r )^2}  \\
& \frac{{\Gamma}_{WW}}{{\Gamma}_{\gamma\gamma}}=\frac{2}{(s^2_w  +c^2_w r )^2}\,.
\end{split}
\ee
It turns out that the branching ratios into $ZZ$ and $W^+W^-$ are potentially larger than the diphoton branching ratio. 
Interestingly, these final states are for the moment not very constrained at the 13~TeV LHC. 
Were the di-photon excess really due to a scalar resonance, it may  certainly be interesting to scrutinize these  channels. This would provide complementary constraints that would readily disentangle between the two effective operators $\phi (Z_{\mu\nu})^2$, $\phi (W_{\mu\nu}^a)^2$ and check the consistency of the description. Such feature comes only from the decays, and is also valid  beyond the case of production by photon fusion.

\subsection{Predictions for Elastic Production}

Finally, maybe the most striking  prediction implied by our effective Lagrangian of Eq.~(\ref{eq:OB_OW})  is that of central exclusive light-by-light scattering. This corresponds to the upper diagram in Fig.~\ref{diagrams}, where protons remain intact and scatter elastically in the forward directions. 
These scattered protons can be potentially detected using Roman pots along the beam pipe. The installation of such forward detectors is being performed at both ATLAS (AFP, \cite{afp}) and CMS (CT-PPS, \cite{ct-pps}).
The full analysis of the search 
for an excess in di-photon events was presented in Ref.~\cite{Fichet:2014uka,Fichet:2013gsa}.
Detecting the intact protons in the final state provides a high precision control of the background, by comparing the mass and rapidity information obtained using the photons in ATLAS/CMS and the intact protons in the forward detectors. Using this method, most of the background, including pileup, is removed. As a result,  only a few events are enough to reach a high statistical significance. In order to obtain a 5 $\sigma$ discovery in this channel, we estimate that
about 6 signal events would be needed after cuts. 

Using a modified version of the  FPMC generator~\cite{fpmc}, and taking into account the acceptance cuts for the forward detectors, we obtain the elastic cross-section
\be
\sigma_{pp\rightarrow \gamma\gamma pp}=  ( 0.23~{\rm fb})\, \left(\frac{5\, {\rm TeV}}{f_{\gamma}}\right)^4 \left(\frac{45~{\rm GeV}}{\Gamma_{\rm tot}}\right) \,.
\label{eq:XS2}
\ee
The result of the fit to the diphoton excess provides directly the value for $\sigma_{pp\rightarrow \gamma\gamma pp}$. In turn, one can deduce the average luminosity needed to reach a 5 $\sigma$ discovery in this channel.
Using the best fit value for $f_{\gamma}$,  we obtain $\sigma_{pp\rightarrow \gamma\gamma pp}= 0.32  $~fb, and the average luminosity for a discovery turns out to be $\sim 21$~fb$^{-1}$. 

\section{UV completions}
\label{UV}

\subsection{General Considerations}
\label{general}

Up to now we have focused on a spin-0 resonance coupled to photons via the effective interaction 
\be
\mathcal L_{\phi\gamma\gamma}=\frac{1}{f_{\gamma}}\phi \, F_{\mu\nu}^2\,.
\label{eff}
\ee
This interaction is non-renormalizable, hence in any UV complete theory one would need to generate it via the exchange of some massive particles. 

If these new particles carried color, as suggested in Refs.~\cite{Harigaya:2015ezk,
Nakai:2015ptz,
Buttazzo:2015txu,
Franceschini:2015kwy,
Angelescu:2015uiz,
Knapen:2015dap,
Ellis:2015oso,
Gupta:2015zzs,
Molinaro:2015cwg,
Higaki:2015jag,
McDermott:2015sck,
Low:2015qep},
an analogous coupling to gluons would be generated which would dominate the photon coupling by  about an order of magnitude, and photon fusion would be subdominant. Hence an unavoidable consequence of a photon-fusion induced resonance would be the presence of uncolored new particles~\footnote{Simultaneously to our preprint Ref.~\cite{Csaki:2015vek} appeared, which pursues very similar ideas.}.
The presence of such particles is suggested by recent models of "neutral naturalness". In fact, the non-detection of colored states usually associated with natural theories of the electroweak scale, such as stops and fermionic top-partners, has prompted the exploration of models in which particles responsible for cutting off the quadratic divergence of the Higgs mass do not carry color. 
These include models of folded supersymmetry \cite{Burdman:2006tz} (in which the new particles are scalars)  or twin Higgs models \cite{Chacko:2005pe} (which predict new color-neutral fermions). In some variants of these models, these fields do however carry electroweak quantum numbers and hence electric charge \cite{Burdman:2006tz,Cai:2008au}.
Moreover, the twin-Higgs models typically feature the presence of a spontaneously broken global symmetry, such that it is tempting to associate the 750 GeV resonance with the Higgs-like excitation of this breaking.
For another interesting theoretically motivated scenario producing large photon couplings wihtout associated gluon couplings see Ref.~\cite{Abel:2016pyc}.
However, independent of this theoretical motivation, it is interesting to make the minimal assumption of the presence of such particles and explore the consequences on their masses and couplings needed to explain the diphoton resonance, as well as their phenomenological implciations at the LHC. For definiteness we assume that these states have spin-1/2, even though a similar analysis could be done for scalars.

\subsection{A Simple Model}
\label{simplemodel}

Let us thus make the very mild assumption that $\mathcal L_{\phi\gamma\gamma}$ is generated via loops of $N$ massive charged fermions, vector-like under the SM. For simplicity we will assume them to be degenerate with mass $m_f$, charge $Q$, and a renormalizable Yukawa coupling to $\phi$
\be
\mathcal L_{\phi\psi\psi}=\lambda\,\phi\,\bar \psi\psi\,.
\ee
The effective coupling above is then given by
\be
\frac{1}{f_{\gamma}}=\alpha \frac{\lambda}{4\pi}Q^2 N \frac{2}{m_\phi} B(\tau)\,,
\label{pert}
\ee
with $\tau\equiv 4m_\psi^2/m_\phi^2$  and
\be
B(\tau)=\sqrt{\tau}\left[1+(1-\tau)\arcsin^2\left (\tau^{-\frac{1}{2}}\right)\right]\,.
\ee
Notice that this coupling becomes complex below the threshold $\tau <1$, i.e.~when the resonance can decay into a pair of fermions. 
However, even in this case, all of our previous analysis is valid provided that one takes the absolute value for the coupling and keeps in mind that Eq.(\ref{eff}) should not be interpreted as an effective Lagrangian. This is in full analogy to the SM, where the coupling of the Higgs boson to photons and gluons is complex due to the presence of the light fermions.
The function $|B(\tau)|$ is of order unity for a wide range of $\tau$, attaining a maximum at $\tau\approx 0.26$ with $B\approx 1.7$.
For large $\tau$ one enters the decoupling regime with $B\sim m_\phi/3m_\psi$.

Fixing $f_{\gamma}\approx 4.6 $ TeV, 
one obtains then 
$
\frac{\lambda}{4\pi}Q^2 N\approx 6
$ (10) for $\tau = 1/4$ (1). 
This points to a large Yukawa coupling,  or sizable charges/multiplicities.  
Observe that once $\lambda\sim 4\pi/\sqrt{N}$, one expects $\mathcal O(1)$ corrections to these numbers due to unsuppressed higher-loop corrections in the Yukawa coupling.

As already pointed out above, a total width of $\Gamma_{\rm tot}\sim 45$ GeV can be saturated by decays into electroweak gauge bosons for $f_B\gg f_W$ in which case one would have to choose $\tau>1$. In the oppposite case, $f_W\ll f_B$, one can obtain a sizable width into fermions by assuming that the decay into fermions is kinematically open, $\tau<1$.
A very interesting feature of this simple model is that in this case one can obtain further constraints on $N$, $\lambda$ and $m_\psi$. The decay width into fermions is given by
\be
\Gamma_{\phi\to \psi\bar\psi}=\frac{N\,\lambda^2}{8\pi} m_\phi(1-\tau)^\frac{3}{2}\,.
\label{Gam}
\ee
In order to avoid that the width becomes too large in our strongly coupled scenario, one can require 
a mild phase space suppression. For example, a combination that works is $\lambda=5$, $m_\psi=360$ GeV, $N=3$, $Q=\frac{5}{2}$. 

In summary, this very simple model of uncolored, charged fermions can both explain the excess 
(via photon fusion) and the width of the observed resonance. 
The model has just 4 free parameters ($Q$, $N$, $\lambda$ and $m_\psi$), and two combinations are already fixed by Eq.~(\ref{pert}) and (\ref{Gam}) above.

It would certainly be very interesting to explore the embedding of this scenario in one of the models motivated in subsection \ref{general}.  For instance, consider the model of Ref.~\cite{Cai:2008au}. It predicts a charged, uncolored, light fermion (triplet under a new gauge group), the "top quirk", coupling with a Yukawa interaction to a scalar (the radial "Higgs" excitation) which would be identified with the 750 GeV resonance. In this  model, however, the charge of the top quirk is relatively small ($Q=\frac{2}{3}$)  such that the diphoton coupling would be suppressed compared to the above example. A more realistic scenario would thus require larger Yukawas or multiplicities. Note however also that a small width remains a possibility for the diphoton resonance, which would allow to explain the diphoton cross section with a smaller diphoton coupling, see Eq.~(\ref{eq:XS2}). A fully realistic model in this context is outside the scope of the present paper and will be left to future research.

\subsection{Further Predictions}

The model described in subsection \ref{simplemodel} also predicts direct production of these vectorlike fermions at the LHC.
First, a generic pair production from non-resonant processes with initial $Z$, $W$, $\gamma$ always exists, and only depends on the EW quantum numbers of the  fermion. Searches focussed on these processes have been investigated in \cite{Kumar:2015tna}, for various quantum numbers, finding a typically mild discovery potential in case of $N=1$. 
 If the vectorlike fermion is light enough, it could also be observed in elastic $\gamma \gamma$ initiated pair production, when both intact protons are tagged. Interestingly, in this case the diagrams are directly proportional to $Q^2N$. 

Finally, if the decay of $\phi$ into fermions is open, a search for resonant pair production via the decay of $\phi$ itself could be attractive. In that case, the vectorlike fermions tend to be emitted back-to-back with  collimated decays. This resonant topology could provide powerful ways to reduce the backgrounds, that may improve the expected sensitivities obtained in \cite{Kumar:2015tna}. 

Therefore, the mass $m_\psi$ and various combinations of the couplings can potentially be measured in direct production and the model could be efficiently tested and constrained.

\section{Details on the Photon Fusion Cross Section}

This final section contains details about the evaluation of the photon fusion cross-section and its various uncertainties.

\subsection{Photon Fusion at 8 and 13 TeV}

In  our baseline scenario, we do not take into account the finite size of the proton. However, including this feature has two important consequences. First, this reduces the cross section by a certain amount, that we parametrised as $r_{\rm fs}$ in Eq.~(\ref{eq:XS}). A rough way of quantifying the impact of the proton size is to let vary a sharp cut on $\xi$, defined to be the fraction of proton momentum carried away by an emitted photon. A cut $\xi<0.15$, for example,  leads to a suppression factor $r_{\rm fs}=0.63$, while a cut $\xi<0.10$ gives a suppression factor 0.5.
Another, more refined way is to use inputs from nuclear physics in order to model the proton form factor. Using  a simple model of the proton as a liquid drop~\cite{daSilveira:2015hha,Dyndal:2014yea,Harland-Lang:2015cta} of a size varying between 0.7 and 1 fm, it turns out that $r_{\rm fs}\approx 0.5$.

The second interesting effect from proton size is that the suppression is substantially different between $8$ and $13$ TeV. 
This can be understood as follows. In order to emit a photon of about 375 GeV for a beam of 4 TeV, the photon 
needs to carry at least about 10\% of the proton energy. This is suppressed if one considers that the 
proton is a composite object. In contrast, at 13 TeV, the photon only needs to carry at least 5.7\% of the proton energy so that the suppression factor is lower.
Taking into account this energy-dependent suppression factor, one finds that the gain in cross section from 8 to 13 TeV can be as high as a factor of 3.9
(using a sharp cut on $\xi<0.15$, while without this effect one would only have a factor of 2.4).
We conclude that this energy-dependent suppression can further improve the consistency between the possible diphoton excess at 13 TeV and the existing limits from 8 TeV data.

\subsection{Uncertainties in the Inelastic Cross Section}

The other source of uncertainty is related to the inelastic processes. 
On one hand, the cleanest possible di-photon production occurs in the case of 
elastic events  where both protons are intact in the final state and two 
photons are measured in the ATLAS/CMS detector. The exclusive di-photon production is dominated by photon-induced processes compared to
 the usual QCD exclusive diffractive production~
\cite{kmr,Fichet:2013gsa,Fichet:2014uka} at high di-photon masses. 
However, since the intact protons are not measured yet for these events in the ATLAS and CMS 
experiments, one  needs to also consider inelastic production of di-photon via photon fusion, which in fact is dominant over the elastic case. 
This has been discussed in detail in Ref.~\cite{szczurek} in the case of $W$ pair 
production and the situation is completely similar in the di-photon case. In Table 1 of Ref.~\cite{szczurek}, the difference contributions to the $WW$ 
production cross section are analysed. The ratio between the total (inelastic-inelastic+inelastic-elastic+elastic-elastic) and the pure elastic contributions is about a factor 20 with a weak dependence on the di-photon mass. The inelastic contributions are taken into account in the cross section of Eq.~(\ref{eq:XS}) using a scaling parameter $r_{\rm inel}$. 
 The inelastic contributions suffer from theoretical uncertainties because the photon PDFs are only partially known, as detailed in Ref.~\cite{szczurek}. In order to take into account these uncertainties, we allow 
 $r_{\rm inel}$ to vary in between 15 and 25.

\section{Conclusions}

We considered the possibility that the slight excess observed at ATLAS and CMS is a spin-0 resonance mostly coupled to electroweak gauge fields. 
This scenario is somehow the most minimal possible, and is described by only one relevant effective coupling, $f_\gamma^{-1}$. 

In this scenario,  the scalar is  mainly produced via a photon fusion mechanism, for which we provide an estimate including the elastic photon fluxes and the photon density function in the proton. 
We observe that taking into account more realistic models of the proton increase the ratio between $8$~TeV and $13$~TeV production rates, which can help  improving the consistency with bounds from run~I.

The coupling to photons  $f_\gamma^{-1} \phi(F_{\mu\nu})^2$ is found to be  $f_\gamma\in[3.8,7.2]$~TeV at 95\%CL.
It turns out that the constraints from the total diphoton event rate and total width are consistent over a large part of parameter space.  This minimal setup automatically provides predictions for the $WW$, $ZZ$ and $Z\gamma$ rates, that can typi\-cally be larger than the diphoton rate. 

In order to get a clean signal with a negligible background, it is important to measure the elastic cross section by tagging both protons. 
Taking into account the efficiencies and acceptance of the forward proton 
detectors, the signal cross section is about  0.32 fb with a negligible background. A luminosity of about 20 fb$^{-1}$ would thus be enough in order to 
obtain a very clean signal at 5$\sigma$ in the di-photon channel.

 
 The scenario we propose admits a very simple UV-completion by assuming the existence of an uncolored vectorlike fermion strongly coupled to the  resonance. 
At the same time, the observed width can be  explained by the resonance decaying into a pair of such  fermions, providing then a prediction for the fermion mass.

\begin{acknowledgments}
We would like to thank Victor Gon\c calves, Cyrille Marquet and Eduardo Pont\'on for illuminating discussions.
G.G and S.F. acknowledge the Funda\c c\~ao de Amparo \`a Pesquisa do Estado de S\~ao Paulo (FAPESP) for financial support. 
\end{acknowledgments}

\bibliography{paper}

\end{document}